\documentclass[prb,preprint]{revtex4-1} 

\usepackage{amsmath}  
\usepackage{amsfonts} 
\usepackage{graphicx} 
\usepackage{tabularx}

\usepackage{array}
\usepackage{makecell}

\usepackage{geometry}
\usepackage{multirow}
\usepackage{booktabs} 

\usepackage{comment}

\begin{document}
\title{The SIREN Program: A Scalable Model for Short-Term Undergraduate Research Experiences at Community Colleges}

\author{Emilie Hein}
\email{heine@smccd.edu} 
\altaffiliation[permanent address: ]{3300 College Drive, San Bruno, CA USA} 
\author{Polin Yadak}
\author{Denise Hum}
\author{Jessica Hurless}
\author{Luis Jibaja Prado}
\author{Susanne Schubert}
\author{Pia Walawalkar}
\author{Marco Wehrfritz}
\author{Aryanna Mendoza}
\affiliation{Department of Physics, Skyline College, San Bruno, CA 94066}

\author{Daria Baitazarova}
\affiliation{California Polytechnic State University, San Luis Obispo, CA 93407}
\author{Yuting Duan}
\affiliation{University of California, San Diego, La Jolla, CA 92093}
\author{Tin Htoo}
\author{Codie Lai}
\affiliation{University of California, Berkeley, Berkeley, CA 94720}
\author{Eslin Villalta}
\affiliation{San Jose State University, San Jose, CA 95192}
\author{Valeria Zarco}
\affiliation{University of California, Davis, Davis, CA 95616}

\date{\today}
\begin{abstract}
Providing meaningful research experiences for undergraduate students is a well-recognized challenge, particularly at community colleges and teaching-focused institutions where resources are limited and faculty time is dedicated to instruction. To address this, the Summer Introduction to Research and  Experimentation in Nuclear physics (SIREN) was developed as a three-week, intensive summer program that engaged 24 students working in six teams on original research projects centered around cosmic detection. Each team was supported by an advisor and a peer mentor, combining structured guidance with near-peer learning.
After developing a range of technical skills, students built cosmic detectors and used them to explore a variety of projects. The program also included multiple workshops on research skills and seminars led by guest speakers, giving students access to broader STEM pathways.
The evaluation of the program was based on advisor and student surveys, as well as peer mentor observations. The findings suggest that the short-term team-based model promoted engagement, collaboration, and skill development while leading to the completion of advanced projects. The advisors highlighted the effectiveness of the program structure and the excitement generated by the direct application of newly acquired skills to research. Data showed that students were able to significantly develop transferable skills through hands-on activities. Post-program surveys indicated that participants experienced an increased sense of belonging in their majors and greater confidence in pursuing careers in STEM. 
\end{abstract}
\maketitle 

\section{Introduction}
Research experiences for undergraduate students are widely recognized as a high-impact practice that enhances students’ technical skills, confidence, and sense of belonging in STEM fields. However, access to these opportunities is often limited, particularly at teaching-focused institutions and institutions serving historically underrepresented populations. The Summer Introduction to Research and Experimentation in Nuclear physics (SIREN) was developed at a community college in a diverse urban region and a designated Hispanic-Serving Institution (HSI), which faces common barriers to providing students with research opportunities, including limited resources, as well as student and faculty time. In 2022, 47.6 percent of full-time students at two-year institutions worked, and 74.2 percent of these working students worked more than twenty hours per week \cite{NCES_2022}. SIREN is a short-term, intensive, team-based undergraduate research experience designed to maximize engagement and learning within a condensed time frame. This paper describes the SIREN model, shares lessons learned from its implementation, and provides a framework for other institutions seeking to adopt similar approaches to broaden access to undergraduate research.

\section{Program Description}
\subsection{Context}
SIREN emerged as a need to offer opportunities that were already existing but limited at the college. An internship program had already been established thanks to the US Department of Energy (DOE) Office of Science (SC) Reaching a New Energy Sciences Workforce (RENEW) program, which was created to offer new opportunities for research by providing traineeships at academic institutions that have historically been underrepresented in the SC portfolio. Yearlong internships in advanced nuclear physics were offered to four to six students per year. To support interns and help them build skills and confidence to tackle complex projects, resources such as technical workshops and networking opportunities were developed. The college is affiliated with the nEXO Collaboration \cite{nexo} and took advantage of connections at a local four-year university and national laboratory to access research topics and guidance for interns. 

The internship program proved successful and revealed a strong student interest in participating in on-campus research opportunities. For example, 30 students applied for the program in Fall 2023, although only six spots were available. Considerable effort and resources had already been invested in developing a framework to support student researchers; however, its reach remained limited. To meet the growing demand and make full use of the existing infrastructure, a shorter program capable of serving a larger number of students was proposed.

\subsection{Program Planning and Structure}
The program was implemented over a three-week period and served 24 undergraduate students, representing a range of STEM disciplines, including Computer Science, Computer Engineering, Electrical Engineering, Mechanical Engineering, Data Science, Biology, Physics, and Mathematics. The participants were organized into six teams of four, which created a collaborative cohort-based research experience while maintaining a manageable mentoring structure. Fourteen of the participants identified as female, reflecting a strong representation of women in the program and further contributing to a learning environment that emphasized inclusion and collaboration. 

The program used a layered mentorship approach that combined faculty guidance with peer support. Each student team was assigned an advisor who provided direction on research goals, technical development, and project milestones. The advising team included faculty from a range of STEM disciplines, namely Chemistry, Mathematics, and Physics, as well as a library faculty member, the STEM Center Program Services Coordinator, and the Dean of STEM (formerly a Communication Studies faculty member). In addition, students with prior research experience served as peer mentors to offer day-to-day support, help troubleshooting technical challenges, and foster collaboration within and across teams. Near-peer mentoring is an evidence-based practice shown to increase students’ sense of belonging, reduce intimidation when seeking help, and enhance engagement in STEM learning environments \cite{Zaniewski_2016}. This dual approach created a supportive learning environment that reinforced skill development, encouraged autonomy, and promoted a sense of community among participants.
\begin{table}[h!]
\centering
\caption{Proof of concept structure of the SIREN program. Students progressed from skill-building workshops to collaborative research projects under the guidance of advisors and peer mentors, culminating in team presentations at a final symposium.}
\label{tab:proofconcept}
\begin{ruledtabular}
\begin{tabular} {lccc}
\textbf{Component} & \textbf{Description} & \textbf{Participants} & \textbf{Outcome} \\
\hline
\textbf{Structure} & \makecell{Six interdisciplinary \\ teams} & \makecell{Four students \\ per team} & \makecell{Collaboration \\ and peer learning} \\
\hline
\textbf{Support Network} & \makecell{Each team guided \\ by one advisor, \\ one peer mentor, \\and additional advisors} & \makecell{Six advisors,\\ six peer mentors,\\ two workshop \\ facilitators} & \makecell{Scaffolded mentorship \\ and accessible \\ research support} \\
\hline
\makecell{\textbf{Phase 1:} \\ \textbf{Instructional}\\ \textbf{components} \\ Weeks 1 and 2} & \makecell{Hands-on \\ technical workshops \\ and essential skills \\sessions} & \makecell{All teams, \\ three teams\\ at a time, \\ led by advisors} & \makecell{Development of \\ technical and research \\ skills} \\
\hline
\makecell{\textbf{Phase 2:} \\ \textbf{Research Projects} \\ Weeks 2 and 3} & \makecell{Team-based \\ investigations using \\ cosmic detectors} & All teams & \makecell{Application of skills \\ to authentic \\ research problems} \\
\hline
\makecell{\textbf{Phase 3:} \\ \textbf{Project Presentations} \\ Week 3} & \makecell{Final symposium \\ with oral presentations} & All teams & \makecell{Communication skills \\ and scientific \\ dissemination}
\end{tabular}
\end{ruledtabular}
\end{table}
The teams had been selected from among 65 applicants. The participants were hired and compensated for three weeks, including two weeks between the Spring 2025 semester and Summer 2025 session, during which they received training in research and technical skills. Students were split into two groups of twelve during most workshops to improve the facilitator-to-student ratio and ensure everyone had hands-on time with the equipment and tools. During the third week, the teams worked more independently to complete their projects, with support from their advisors and peer mentors, and gave final presentations on the last day of the program. The proof of concept structure is described in Table \ref{tab:proofconcept}.

\subsection{Instructional Components}
In addition to supporting teams of students, advisors facilitated a number of workshops. Some focused on building technical skills, while others were designed to strengthen essential skills students need to become well-rounded STEM professionals.

\subsubsection{Developing technical skills}
In order for students to be able to define and get started with their research projects, they needed to develop a specific set of technical skills. To do so, they completed the following workshops. 
\begin{itemize}
    \item \textbf{Physics workshops}: Students were introduced to the basics of nuclear physics, radiation, and cosmic detection. This was a pre-requisite to understanding how a cosmic detector operates and possible uses. 
    \item \textbf{Hands-on technical workshops}: Students gained advanced soldering skills, and were introduced to basics of electronics, microcontrollers, 3D design and printing. They applied their newly acquired skills to building a CosmicWatch \cite{Axani_2018}. They also learned how to use phyphox\cite{Staacks_2018}, a smartphone-based physics experimental platform. Using built-in sensors, students conducted a speed of sound experiment, collecting and analyzing real data to explore how to identify and propagate error in their measurements.
    \item \textbf{Python for data analysis}: Using cosmic detection examples and real data collected from a high-altitude balloon flight with local partners, students learned the basics of Python programming for data visualization and analysis.
    \item \textbf{AI in scientific research}: Students learned to use generative AI tools to refine research questions, craft more effective prompts, and locate scholarly literature.
\end{itemize}

\subsubsection{Developing research skills and beyond}
In addition to technical training, the program incorporated a series of activities aimed at developing students’ professional and personal skills that could enhance their research experience. These included team building exercises, the development of a portfolio to document progress, and targeted workshops on communication and presentation skills, including crafting an effective elevator pitch. Students also received structured guidance in strengthening their research skills, while broader sessions encouraged them to think beyond transfer, exploring potential career paths, internships, and graduate school opportunities. To address challenges often encountered by students in STEM, the program also created space for open discussions around imposter syndrome, helping participants build confidence and resilience.

\subsubsection{Guest speakers}
The program also featured eight guest speakers who shared their personal journeys, professional insights, and advice, further connecting students to the real world impact and diversity of careers in STEM research. Speakers included recent alumni who had transferred to four-year universities after completing the internship program at the college, as well as past and current research mentors with expertise in particle and nuclear physics. They spanned multiple career stages, ranging from early-career scientists to senior researchers, and included professionals affiliated with both academic institutions and national laboratories. These sessions usually took place over lunch and provided students with role models, broadened their awareness of career paths, and helped them envision their own trajectories within and beyond academia.

\subsection{Research projects}

After completing the training components of the program, each team had a set of CosmicWatch detectors they had built themselves to use in their research projects. By this stage, students had begun brainstorming potential research topics and had prepared extensive annotated bibliographies to survey prior work in the field and refine their ideas. Their progress and research process were documented in electronic portfolios. Project topics varied widely, reflecting both the personal interests of the students and the technical skills they sought to further develop.

One area of focus was the application of CosmicWatch detectors to high-altitude balloon experiments, where students investigated how flight conditions had affected data collection and proposed the following improvements for future launches. 
\begin{itemize}
\item Stabilized Platform for High-Altitude Cosmic Ray Detection
\item Spin-Dependent Sensitivity: Testing CosmicWatch Response to Rotational Motion
\item A Rotating Dual-Detector CosmicWatch System for Reducing False Coincidences and Evaluating Detection Efficiency
\end{itemize}
Other teams explored applications of muography, examining how detectors could be used to probe material properties or image structures.
\begin{itemize}
\item Exploring Muon Flux Angular Dependence and Material Attenuation using a Single CosmicWatch Detector
\item Effectiveness of Liquid-Based Shields in Muon Flux
\item Muography: Create a Digital Image without Measuring Density
\end{itemize}
On the final day of the program, all teams delivered oral presentations and showcased demonstrations of their prototypes. Students were encouraged to invite guests, resulting in the attendance of several family members and supporters. The event served not only as a celebration of the students’ hard work but also as an opportunity to reflect on their growth and achievements over the course of the program.

\subsection{Cost Estimate}
All participants received compensation for their time in the program. Students were paid for their three weeks of engagement, peer mentors received compensation for the same number of hours, and faculty advisors were supported for the time required to design and deliver workshops as well as mentor the student teams.

Each student was assigned a CosmicWatch detector, with components purchased in advance and organized into kits by peer mentors to streamline assembly. Additional supplies for workshops and research activities were also procured.

To foster community and facilitate interaction with guest speakers, lunch was provided for all participants during the first two weeks of the program and on the final presentation day.

\section{Evaluation Methods}

The SIREN program was assessed using a mixed-methods evaluation framework that captured outcomes at the student, mentor/advisor, and programmatic levels. 

Pre- and post-program surveys were administered to students to assess their growth across key learning outcomes and to evaluate changes in their sense of belonging in STEM. Observations were also used to measure growth in research skills, technical competencies (e.g., coding, soldering, data analysis), and teamwork.

Advisors completed feedback surveys and participated in debrief discussions to assess workload, professional development, and sustainability of their involvement. Peer mentor evaluations focused on their experiences supporting student learning, balancing guidance with independence, and developing leadership and communication skills.

Evaluation also considered the resources, logistics, and structures that enabled the program’s success. Cost tracking, workshop effectiveness, and schedule feedback were analyzed to identify strengths and areas for improvement. This was particularly interesting in order to evaluate whether or not the program was efficient, adaptable to different institutional contexts, and scalable through intentional use of peer mentors, cross-campus faculty, and modest infrastructure.

Together, these evaluation components provided a holistic understanding of how SIREN functioned as both an educational and organizational model, highlighting its contributions to student development, mentor growth, and institutional capacity.

\section{Results and Findings}

\subsection{Quantitative Feedback}
The SIREN program yielded meaningful outcomes for students, mentors, and the institution. Participants (n = 21) completed identical pre- and post-surveys across 12 skill domains rated on a 0–5 scale. Here, $d$ denotes Cohen’s $d$ (standardized mean difference, using the pooled standard deviation (SD)). As shown in Table \ref{tab:preproscores}, mean scores ($\Delta$M) increased substantially across all items, with average gains of +2.0 points. Improvements were especially pronounced in nuclear physics ($\Delta$M = 2.9, \textit{d} = 2.51), phyphox ($\Delta$M = 2.6, \textit{d} = 2.52), and basic electronics ($\Delta$M = 2.7, \textit{d} = 2.43). Smaller but still significant gains occurred in areas where students already had moderate familiarity, such as graphing ($\Delta$M = 1.2, \textit{d} = 1.10), scientific poster ($\Delta$M = 1.7, \textit{d} = 1.12), and presentation skills ($\Delta$M = 1.6, \textit{d} = 1.23).
All effect sizes were large (Cohen’s \textit{d} range = 1.10 – 2.52), indicating the workshop produced strong practical learning benefits. 
\begin{table}[ht]
\centering
\caption{Pre- and Post-Program Scores with Cohen’s $d$}
\label{tab:preproscores}
\setlength{\tabcolsep}{8pt}
\renewcommand{\arraystretch}{1.15}
\begin{tabular}{@{}lcccc@{}}
\hline\hline
\textbf{Item} & \textbf{Pre M (SD)} & \textbf{Post M (SD)} & \textbf{$\Delta M$ (Post--Pre)} & \textbf{Cohen’s $d$} \\
\hline
Basic Electronics & 2.2 (1.1) & 4.9 (0.3) & 2.7 & 2.43 \\
Soldering         & 2.4 (1.4) & 5.0 (0.2) & 2.6 & 1.89 \\
Nuclear Physics   & 1.6 (0.8) & 4.5 (0.7) & 2.9 & 2.51 \\
Using Arduino     & 2.0 (1.2) & 4.1 (1.0) & 2.1 & 1.70 \\
Phyphox           & 1.4 (0.8) & 4.0 (0.9) & 2.6 & 2.52 \\
Python            & 2.0 (1.1) & 4.1 (0.9) & 2.2 & 1.49 \\
Graphing          & 2.9 (1.1) & 4.1 (0.9) & 1.2 & 1.10 \\
3D Printing       & 2.5 (1.5) & 3.9 (1.0) & 1.3 & 0.99 \\
AI                & 2.5 (1.4) & 4.6 (0.7) & 2.0 & 1.58 \\
Research          & 2.4 (1.2) & 4.7 (0.4) & 2.3 & 1.84 \\
Scientific Poster & 2.6 (1.3) & 4.4 (0.9) & 1.7 & 1.12 \\
Presentation      & 2.9 (1.2) & 4.5 (0.8) & 1.6 & 1.23 \\
\hline\hline
\end{tabular}
\end{table}
In addition to the pre-post survey, participants completed 15 post-only items (Likert 1–5) assessing their perceptions of the program. Across all questions, mean score were significantly higher than the neutral midpoint of 3 (p $< .001$), indicating strong endorsement of the program’s effectiveness (Tables \ref{tab:survey_results_1} and \ref{tab:survey_results_2}). 
Students particularly valued the hands-on learning components: all participants rated “I enjoyed the hands-on activities more than traditional classroom learning.” At the maximum score (M = 5.0, SD = 0.0). They also reported high engagement (M = 49), creativity and problem-solving gains (M = 4.9), and confidence-building from final presentations (M = 4.5). 
Feedback on program impact was likewise positive. Students felt more motivated to pursue STEM-related opportunities (M = 4.9), learned about different STEM career pathways (M = 4.6), and expressed high satisfaction overall (M = 4.8). Importantly, students also reported confidence in mentoring future cohorts (M = 4.6), suggesting sustainability and peer-leadership potential.
\subsection{Qualitative Feedback}
Analysis of open-ended responses yielded four major themes: hands-on learning, collaboration and mentorship, program structure and organization, and personal growth.
\begin{itemize}
\item Hands-on Learning: Students emphasized soldering, electronics, and 3D printing as highlights. One noted, “What I enjoyed most was applying what I had learned in class to a real research project… seeing the system work was incredibly exciting.”
\item Collaboration and Mentorship: Many valued teamwork and supportive mentoring. As one student shared, “I enjoyed that I got to share my ideas with my teammates and we each respected each other’s strengths.” Another commented, “What I enjoyed most was getting hands-on experience and collaborating with others. I learned more when I was part of a team.”
\item Program Structure and Organization: Suggested improvements included extending program length, adjusting the timing of guest speakers, and providing more detailed instruction in workshops. For example, “Python workshops were fast paced and I think more explanation on the graphing and visualizing would be helpful.”
\item Personal Growth and Impact: Several described the experience as transformative. One participant wrote, “I was hesitant about pursuing computer science when I transferred, but this program helped me try different fields and realize how much I want to stay in STEM, even if it seems daunting.”
\end{itemize}
\begin{table}[ht]
\centering
\caption{Survey Results: Program Activities Post-Program Reflections}
\label{tab:survey_results_1}
\setlength{\tabcolsep}{7pt}       
\renewcommand{\arraystretch}{1.2}
\begin{tabular}{@{}p{10cm}cccc@{}}
\hline\hline
\textbf{Survey Questions} & \textbf{M} & \textbf{SD} & \it{t} & \textbf{p} \\
\hline
The CosmicWatch activity helped me connect science to real-world tools. & 4.7 & 0.6 & 14.4 & $<.001$ \\
The guest speakers were informative and inspiring. & 4.8 & 0.5 & 16.6 & $<.001$ \\
I enjoyed the hands-on activities more than traditional classroom learning. & 5.0 & 0.0 & -- & -- \\
I felt engaged and interested throughout the program. & 4.9 & 0.3 & 29.7 & $<.001$ \\
The program encouraged me to think creatively and solve problems. & 4.9 & 0.3 & 29.7 & $<.001$ \\
I had enough support from advisors and mentors when I needed help. & 4.9 & 0.4 & 24.3 & $<.001$ \\
Preparing for the final presentation helped me reflect on what I learned. & 4.7 & 0.6 & 11.9 & $<.001$ \\
Overall, I am satisfied with my experience in the program. & 4.8 & 0.4 & 21.1 & $<.001$ \\
I would recommend this program to other students. & 5.0 & 0.2 & 42.0 & $<.001$ \\
\hline\hline
\end{tabular}
\end{table}
\begin{table}[ht]
\centering
\caption{Survey Results: Individual Growth Post-Program Reflections}
\label{tab:survey_results_2}
\setlength{\tabcolsep}{7pt}       
\renewcommand{\arraystretch}{1.2}
\begin{tabular}{@{}p{10cm}cccc@{}}
\hline\hline
\textbf{Survey Questions} & \textbf{M} & \textbf{SD} & \it{t} & \textbf{p} \\
\hline
I better understand communication styles and how to communicate effectively when working as part of a team. & 4.7 & 0.8 & 9.9 & $<.001$ \\
I learned about different STEM career pathways during the program. & 4.6 & 0.7 & 15.2 & $<.001$ \\
This program helped me understand what I might want to study or do in the future. & 4.6 & 0.7 & 10.9 & $<.001$ \\
I feel more motivated to pursue STEM-related opportunities after this experience. & 4.9 & 0.3 & 29.7 & $<.001$ \\
Presenting my work boosted my confidence in communicating technical ideas. & 4.5 & 0.8 & 8.6 & $<.001$ \\
I feel confident about attending a similar program in the future as a student mentor. & 4.6 & 0.6 & 12.9 & $<.001$ \\
\hline\hline
\end{tabular}
\end{table}
Overall, students demonstrated high levels of engagement and reported growth in confidence and technical skills, with many participants expressing that the program helped clarify their academic and career goals. The impact on students exceeded expectations, especially in the way it increased their sense of belonging in STEM, at the college and beyond. One student described the program as “life-changing” and credited it with giving them the courage to speak up more and contribute meaningfully to team projects, despite initial concerns about language barriers. Other participants emphasized how the cohort structure, mentorship, and workshops contributed to feeling part of the larger STEM community at the college. “The program itself and the amount of engagement I’ve had ever since the first day SIREN started… This has not only boosted my confidence to present in a group of people but taught me the importance of hands-on learning and resilience itself.” Another student added, “I really felt that I belonged in the program, developed a lot of professional research-related skills, and became inspired to pursue a career in research.”

\subsection{Advisor and Mentor Outcomes} 

Faculty advisors and peer mentors reported that the condensed program format was sustainable and effective. Advisors noted that while the time commitment was concentrated, it allowed them to provide meaningful mentorship without the ongoing workload of a semester-long project. Peer mentors described the experience as an opportunity to grow professionally, gain leadership skills, and support students in a way that reinforced their own technical learning. Students themselves highlighted the importance of this mentorship, with one noting, “The best part of the program was the student mentors. They were really helpful and supported the learning process without doing things for us, so we still had the opportunity to grow.”

\subsection{Program Outcomes} 
Program outcomes included the successful assembly and operation of a large numbers of CosmicWatch detectors, which can be used in physics laboratory setting, the preparation of prototypes for high-altitude balloon payloads, and the creation of a collaborative, research-focused environment. Students found it particularly rewarding to apply classroom knowledge to real projects: “It was incredibly exciting to assemble and operate the CosmicWatch detectors, write code to run the experiment, and actually see the system work. Being able to connect theoretical concepts with hands-on work made the entire experience deeply meaningful.”

Both students and mentors noted challenges, including the complexity of data analysis, detector calibration, and the fast-paced timeline. Some participants recommended mid-program check-ins or additional time for project development to enhance collaboration across groups. 

These challenges, however, were generally framed as learning opportunities that strengthened problem-solving skills and resilience.
Several elements contributed to the program’s success: a short-term but structured format, clearly defined deliverables such as annotated bibliographies and final presentations, and the integration of peer mentors as near-peer role models. Students valued the balance between technical training and exploratory research, as well as the opportunity to engage directly with faculty and external speakers.

Unlike research experiences for undergraduates (REUs), which typically span an entire summer, or course-based undergraduate research experiences (CUREs), which unfold across a semester, this model offers a short, intensive format that is highly accessible in community college settings. It requires fewer resources than most REUs while providing more structure and mentoring than typical CUREs.

Key takeaways include the importance of establishing clear project goals early, ensuring a balance between structured training and open-ended inquiry, and leveraging peer mentors as an essential layer of support.

Because the program utilized low-cost equipment, modular workshops, and a peer-mentorship framework, it is both adaptable and replicable. As one student summarized, “This program is a really good résumé builder and has given me access to many opportunities.” These features suggest that the SIREN model could be successfully adopted by other institutions with limited resources, broadening access to authentic research experiences.

\section{Conclusion and Future Work}

The SIREN program demonstrates that short-term, resource-efficient undergraduate research experiences can provide meaningful engagement, skill development, and a greater sense of belonging for students in physics and related STEM fields. By combining technical workshops, peer mentorship, and team-based projects, the program offered an accessible framework that balanced structured training with authentic inquiry. Student feedback highlighted increases in confidence, collaboration, and research skills, while advisors noted the sustainability of the model and its alignment with faculty workload. 

A distinctive feature of the program was its multidisciplinary advising team, which drew faculty and staff from across the college. In addition to STEM faculty, contributors included library faculty, program coordinators, and administrators. Their involvement demonstrated how institutions can leverage the diverse strengths of their community, including individuals outside traditional STEM disciplines, to create richer learning environments and more sustainable support structures.

Future iterations of the program will focus on expanding the scope of projects, opportunities for longer-term or follow-up projects, and additional engagement with external speakers and collaborators. These enhancements aim to deepen the impact of the program while preserving accessibility to research experiences.

We present SIREN as a replicable framework for institutions seeking to expand undergraduate research opportunities without the extensive resources required by traditional REUs. Importantly, this model demonstrates that meaningful research engagement can be achieved in community colleges and other teaching-focused institutions through intentional design, modest investment, and the integration of peer mentorship.

Expanding access to undergraduate research matters because it helps diversify the STEM pipeline and equips students with the skills and confidence to pursue advanced study and careers. The results of this program suggest that the SIREN model can serve as a practical, scalable approach that other colleges can adopt or adapt to meet the needs of their students and faculty.

\begin{acknowledgments}

The authors gratefully acknowledge the Skyline College community for its strong support of the SIREN program, particularly the STEM Division, whose collaboration and commitment made this work possible. We also thank the guest speakers, nEXO collaborators, and our partners at universities and national laboratories for sharing their expertise and contributing to a rich research environment for students.

This work was supported by the U.S. Department of Energy Office of Science (Office of Nuclear Physics) under Award Number DE-SC0024677. Additional institutional support was provided by Skyline College’s Office of Planning, Research, Innovation, and Effectiveness (PRIE), which also reviewed and approved this study. 

\end{acknowledgments}

\section{Author Declarations}

\subsection{Conflict of Interest}

The authors declare no conflict of interest.

\subsection{Ethics Approval Statement}

This study was reviewed and approved by the [institution] Office of Planning, Research, Innovation, and Effectiveness (PRIE) and was determined to be exempt under 45 CFR 46.104(d)(1) for research conducted in established or commonly accepted educational settings involving normal educational practices. Participation was voluntary, and all data were collected anonymously. Informed consent was obtained by survey completion, with participants informed that their responses would remain confidential and that no identifying information would be collected.


\begin{thebibliography}{99}

\bibitem{NCES_2022} National Center of Education Statistics. \url{https://nces.ed.gov/programs/digest/d23/tables/dt23_503.40.asp}

\bibitem{nexo} G. Adhikari et al,``{nEXO}: neutrinoless double beta decay search beyond $10^{28}$ year half-life sensitivity'' J. Phys. G: Nucl. Part. Phys. \textbf{49}, 015104 (2022). \url{https://doi.org/10.1088/1361-6471/ac3631}

\bibitem{Zaniewski_2016}
A.M. Zaniewski and D. Reinholz, ``Increasing STEM success: a near-peer mentoring program in the physical sciences.'' IJ STEM Ed \textbf{3}, 14 (2016). \url{https://doi.org/10.1186/s40594-016-0043-2}

\bibitem{Axani_2018}
S.N. Axani, K. Frankiewicz, and J.M. Conrad, ``The {CosmicWatch Desktop Muon Detector}: a self-contained, pocket sized particle detector'' JINST \textbf{13} P03019 (2018). \url{http://dx.doi.org/10.1088/1748-0221/13/03/P03019}
 
\bibitem{Staacks_2018} S. Staacks, S. Hütz, H. Heinke, and C. Stampfer, ``Advanced tools for smartphone-based experiments: phyphox'' Phys. Educ. \textbf{53} 045009 (2018).  \url{https://doi.org/10.1088/1361-6552/aac05e}.

\end{thebibliography}
\end{document}